# Human-GDPR Interaction: Practical Experiences of Accessing Personal Data

Alex Bowyer
Open Lab, Newcastle University, Newcastle upon Tyne, United Kingdom
a.bowyer2@newcastle.ac.uk

Jack Holt
Open Lab, Newcastle University, Newcastle upon Tyne, United Kingdom
j.holt3@newcastle.ac.uk

Dr. Josephine Go Jefferies
Newcastle University Business School, Newcastle University, Newcastle upon Tyne, United Kingdom
josephine.go-jefferies@newcastle.ac.uk

Prof. Rob Wilson, PhD
Faculty of Business and Law, Northumbria University, Newcastle upon Tyne, United Kingdom
rob.wilson@northumbria.ac.uk

Prof. David Kirk
Open Lab, Newcastle University, Newcastle upon Tyne, United Kingdom
david.kirk@newcastle.ac.uk

Dr.-Ing. Jan David Smeddinck
Open Lab, Newcastle University, Newcastle upon Tyne, United Kingdom
jan.smeddinck@newcastle.ac.uk

## ABSTRACT

In our data-centric world, most services rely on collecting and using personal data. The EU's General Data Protection Regulation (GDPR) aims to enhance individuals' control over their data, but its practical impact is not well understood. We present a 10-participant study, where each participant filed 4-5 data access requests. Through interviews accompanying these requests and discussions scrutinising returned data, it appears that GDPR falls short of its goals due to non-compliance and low-quality responses. Participants found their hopes to understand providers' data practices or harness their own data unmet. This causes increased distrust without any subjective improvement in power, although more transparent providers do earn greater trust. We propose designing more effective, data-inclusive and open policies and data access systems to improve both customer relations and individual agency, and also that wider public use of GDPR rights could help with delivering accountability and motivating providers to improve data practices.

## 1 INTRODUCTION

The modern world is data-centric; our everyday lives are entwined with the digital. Organisations that provide services collect our personal data, often without *fully informed* consent, to facilitate algorithmic decision-making. Data has become a commodity, exploited and traded for commercial advantage and for behavioural insights that serve advertisers before users. Personal data has become providers' property, and the individuals concerned are unable to easily see, access or understand how it is being used. These issues affect the experience of using digital services, and therefore good *human-data interaction* (HDI) [84] should be a core matter of concern for HCI and UX professionals [85].

There is a power imbalance over personal data [48–51]; it is currently scattered and trapped beyond individual reach. In 2018, the European Union's General Data Protection Regulation (GDPR) [36] came into force as an attempt to rebalance power by granting individuals rights to access their data and have its usage explained. In the four years since, similar policies have been introduced around the world [52], and there has been a visible impact upon providers, who are required to respond to personal data access requests. While the GDPR has become a valuable tool for transparency that has been examined and harnessed by various researchers, few have done so from a user-centric, HDI [85] perspective. Our research seeks to go deeper than prior work [4] by examining the human experience of exercising one's GDPR rights: exploring compliance, response quality, individual attitudes to data-holding organisations, and the impact upon the service relationship. We aim to identify challenges, help individuals inform their choices around data and explore how data policies and practices could be redesigned.

We present qualitative and quantitative findings from an 18-month study. Through analysis of 31 interview transcripts and numerous sketches and spreadsheets, we derived insights for policymakers, data-holding companies and individuals. We find the GDPR's aim to provide individuals with control over—and value from—their data is hampered by non-compliance and poor-quality data responses, and that both data holders and individuals stand to gain from improved data access and transparency. Informed by our findings, we recommend specific design approaches that can inform the different parties involved.

Our key contributions include:

- A detailed account of the extent to which service providers complied with GDPR requests, including types of data received, perceived completeness, accuracy and usability.
- A summary of the hopes, plans and imagined uses of personal data that motivate GDPR requests.
- Thematic findings showing data holders were seen to be evasive and that individuals struggle to make sense or use of returned data.
- Evidence of a detrimental impact upon individuals' trust in data-holding organisations as a result of scrutiny of privacy policies and GDPR responses.
- Sociotechnical design insights to redesign the GDPR for greater effectiveness.
- Advice for companies on restructuring their users' relationship with personal data to improve trust.
- Advice for individuals on how to exploit GDPR rights for positive impact.

## 2 BACKGROUND

We begin by outlining the GDPR legislation and its origins, then we review prior research exploring GDPR, and explain why further human-centred research is needed, building upon recent HDI research and innovation around personal data ecosystems.

### 2.1 GDPR: Legislation Seeking to Empower Individuals in a Data-Centric World

The widescale adoption of personal computers [115] and smartphones [22, 41], combined with the advent of cloud-computing [68], have led to ubiquitous storage of personal data [49, 60] by commercial and public sector service providers. Data-centric companies dominate almost every sector [23, 71]. We live *digital lives* [14, 130] and personal data collection is inevitable. With the rise of data-driven decision making, data is a valuable resource to be mined and exploited at scale [19, 35, 89, 102, 113].

Unfortunately, people have minimal awareness of—or access to—their data. Large-scale data-centric systems that drive modern life largely function as opaque *'data traps'* [1] and data collection is often unwitting [101]. The World Economic Forum's 'Rethinking Personal Data' project recognised the critical role data now holds, noting that "*an asymmetry of power exists [. . .] created by an imbalance in the amount of information about individuals held by industry and governments, and the lack of knowledge and ability of the same individuals to control the use of that information*" [48–51]. Data becomes a proxy for direct involvement in decision-making [18] and without effective access to data [44] people are disempowered, lacking agency and control over data held about them [31, 84].

Since the 1970s laws have aimed to protect individuals' data and data rights [15, 29, 91, 116, 128], although they have been '*almost useless in limiting the growth of surveillance*' [81]. The EU's GDPR [36], through the execution of designed-to-hurt fines [21, 27, 66, 70, 95] for non-compliance, has begun to have some impact [7], giving at least 513 million people[1] rights to timely data access, explanation, erasure and correction [57]. It is regarded as a landmark piece of legislation and a strong template for individual data protection, having inspired similar legislation such as California's CCPA [67] and others in India, Japan, Turkey and beyond. Data protection has become particularly important and gained increased global public awareness due to the Snowden revelations [39], Cambridge Analytica scandal [129] and the COVID-19 pandemic [45, 90].

Since the GDPR's launch in May 2018, many large B2C companies have developed 'privacy hubs' or improved privacy policies where almost anyone[2] can learn how their personal data is handled or easily download copies of it [117–120]. Research is needed to study the effectiveness of these measures for users. Does compelling data holders to create such offerings and respond to access requests enable GDPR to succeed in its goal to '*enhance the data protection rights of individuals*' [30] and to give people '*control over their personal data*' [100] and redress the power imbalance between data holders and individuals[3]?

### 2.2 Current GDPR Research and its Limitations

Since 2018, the adoption of the GDPR has opened up new possibilities for research [28]. For example, the ability to obtain one's data records from organisations affords the general public and the research community insights into organisational processes, and Ausloos and Veale [8, 10] outline an approach for such research and discuss ethical and methodological considerations.

Other prior studies focusing on the GDPR process include (findings in brackets):

- understanding data holders' compliance with legislation [6, 9] (generally poor)
- evaluating data portability [109] (inadequate) and '*privacy by design*' [106] (largely absent)
- assessing GDPR's effectiveness in improving:
  - explainability [45] (challenging to provide),
  - fairness [63] (ill-defined and challenging),
  - consent [53] (insufficiently considered),
  - transparency [97] (approaches are suggested)

---

[1] the population of the European Union and the UK, the jurisdictions in which the GDPR legally applies [37]

[2] In practice these rights are available globally, not just in the EU and UK, as international companies rarely limit GDPR data access mechanisms such as download dashboards or e-mail request handling by geography.

[3] The GDPR uses the following terms:
'*data controller*' (to describe the organisation that determines the purpose and means of data collection,
'*data processor*' (to describe organisations that act with data on behalf of a data controller),
'*data protection authority [DPA]*' (to described the authority in each country responsible for enforcement) and
'*data subject*' (to describe the identified or identifiable natural person to which the stored information pertains).
For simplicity, throughout this paper we use the term '*data holder*' to encompass both data controllers and processors, '*individual*' to refer to the data subject, and '*DPA*' to refer to the data protection authority.

- and in reducing data breach risks [43] (risk assessments should be conducted).

Potential negative impacts have also been considered: the threat to privacy [20, 77] (through insufficient ID verification) and potential impediments to health research [26] (by creating extra burdens).

GDPR impacts legal, societal and technology domains, yet, research the individual experience of the GDPR is scant. Alizadeh *et al.* did conduct a study with 13 users of a German loyalty programme and interviewed them before, during and after they made GDPR data requests [4], finding better responses and GDPR education were needed. However, this study was limited in breadth (targeting only one provider) and depth (returned data was discussed at the high level of '*were your expectations met?*' and potential data uses were not examined). The impact upon the participants' relationship with the provider was not explored. Recent work [20, 42, 114] has established that openness and transparency around data handling are key to services establishing users' trust; in a commercial context this impacts customer satisfaction and business success.

Research is needed to understand the *experience* people have when using the GDPR because to date, companies' GDPR processes have been designed to comply with litigation rather than focusing on user needs or desires [3, 78, 111]. Insights into '*the human experience of the GDPR*' could inform the design of improvements to digital GDPR mechanisms and help identify potential policy improvements.

## 2.3 Human-Data Interaction: Towards a Human-centric Personal Data Ecosystem

In 2017, the average American Internet user had 150 online accounts with different providers [24]. Data for the UK show the number of service and supply relationships each individual has to manage increasing from around 45 in 1997 to around 250 in 2020 [47]. As the amount of personal data per capita has increased, so has the need for individuals to be able to manage it. In the 1990s, this was considered through the *personal information management* (PIM) lens of giving people '*a place for their personal data*' [62] to facilitate easy filing-and-retrieval to improve task efficiency and personal productivity [5]. Subsequently there has been growing recognition that this problem needs to be tackled at a 'whole life' level; our data exists across devices and across providers [2, 88, 107]. Through the fields of *personal informatics* [72, 73] and the *quantified self* [65], researchers and hobbyists have explored the practicalities of *collecting* and *integrating* data about oneself so that the individual is able to *reflect* upon it, gain *insights* and take informed *action* [73, 74, 93]. In the wider sociotechnical context, where data is held by multiple service providers offering limited use and access, the challenge of drawing meaning from one's data increases substantially. Data cannot be moved freely [16] so integrating our data becomes onerous [31]. Managing personal information has transformed from 'arranging your bookshelves' into a multi-party negotiation over representation, ownership, access and consent [101] that people lack the skills, tools and time to manage. Data is a shared resource, often with multiple users. A few researchers have begun to look at people's interactions with data in this context [46, 92, 112] but not considering access to data held by service providers.

Our review of prior research suggests a human-centric approach to personal data is needed, placing individuals at the centre, as controllers and overseers of their own personal data ecosystem [88, 99]. This emerging area of research and innovation [13, 32, 86, 87, 99, 121, 131] provides a frame to evaluate the human experience of – and interaction with – the GDPR. Our research questions are: Given individuals' diminished agency and control over their data [31, 110], do the GDPR's access rights provide the *effective access* [44] people need? Does the GDPR help people to achieve *legibility*, *agency* and *negotiability*, which are needed for effective HDI [85]? Can people acquire a functional relationship *with* their personal data [31]?

Our study addresses these research gaps by documenting the experience of exercising one's GDPR rights and assessing how well the process meets individuals' needs and expectations to uncover problems and possible solutions that could address them.

## 3 STUDY SETUP

We advertised our study as an opportunity to receive GDPR coaching to obtain and explore one's own held data. Using convenience sampling we recruited 11 participants through our connections at Newcastle University, aged 20-40 years, self-identifying as 5 women and 6 men. One woman was French and one Chinese. Everyone else was ethnically British. Time was compensated for with £20 shopping vouchers.

We conducted 31 qualitative interviews in total (each participant was interviewed three times; see Figure 1) between December 2019 and April 2020. The scope and purpose of each interview was as follows:

- **Interview 1: Sensitisation, Life Exploration and Company Selection** (1 hour, in person). Participants were sensitised to GDPR through a discursive tour of a poster display on GDPR rights, potential data-holding organisations, datatypes and potential uses for GDPR-obtained data. Baseline data was collected on participants' hopes and motivations, current understanding of personal data, data access, data control, and power as it relates to data. Using a sketch interviewing [54] technique, participants mapped out their 'data lives' (e.g. Figure 2), annotating organisations that they have relationships with, types of data those companies might hold, and feelings about such data use and storage by each holder. Each participant selected 4-5 target companies.
- **Interview 2: Privacy Policy Reviewing, Goal Setting and GDPR Request Initiation** (1 hour, in person). To stimulate reflective thinking and measure impacts, participants were asked to discuss and score their initial feelings of trust and power with each company (cf. 4.4). Participants then viewed key sections of privacy policies on screen with the researcher, to identify each company's promises regarding personal data. Afterwards, participants initiated email GDPR requests to each company, which had been prepared using a tried-and-tested template generated by personaldata.io [108]. For P10 & P11, interview 2 took place over Zoom due to the COVID-19 pandemic.
- **Interview 3: Detailed GDPR Response Review** (2 hours, online video call). Having allowed the legally-mandated 30

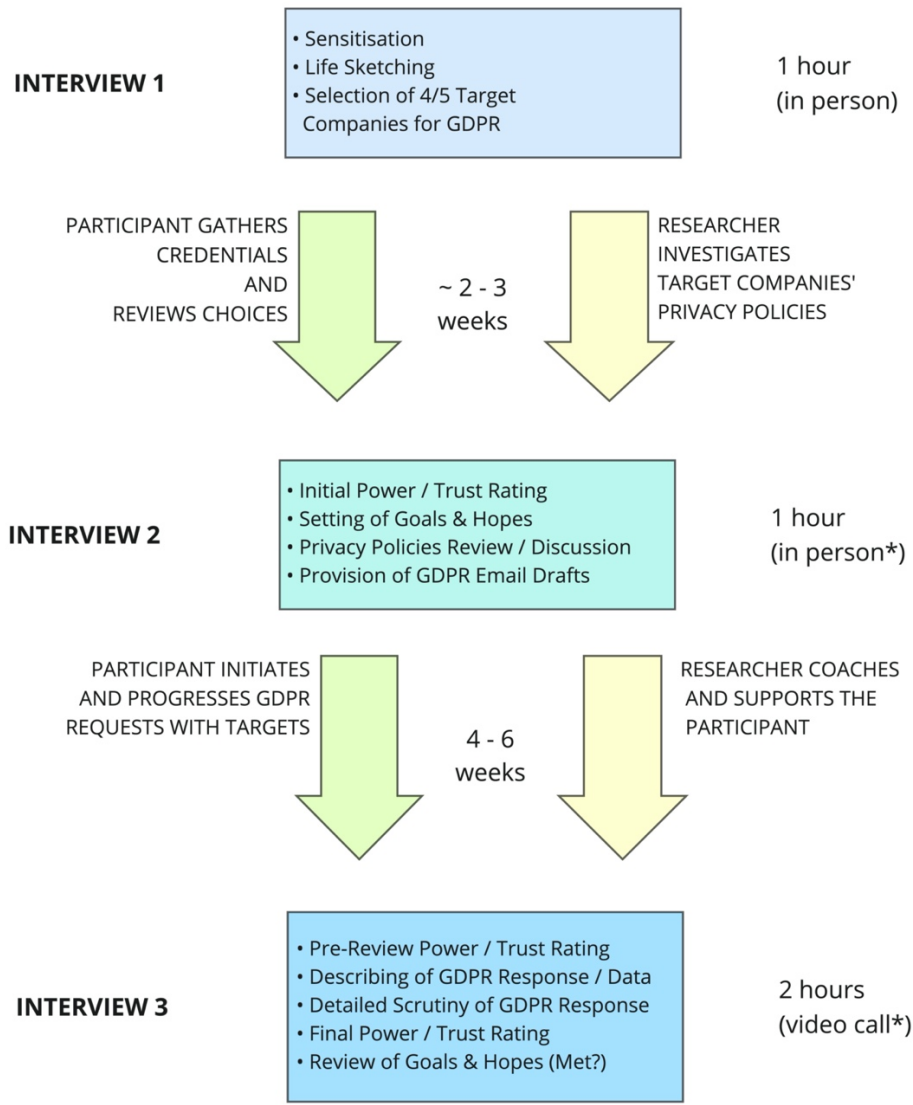

**INTERVIEW 1**
- Sensitisation
- Life Sketching
- Selection of 4/5 Target Companies for GDPR

1 hour
(in person)

PARTICIPANT GATHERS CREDENTIALS AND REVIEWS CHOICES

~ 2 - 3 weeks

RESEARCHER INVESTIGATES TARGET COMPANIES' PRIVACY POLICIES

**INTERVIEW 2**
- Initial Power / Trust Rating
- Setting of Goals & Hopes
- Privacy Policies Review / Discussion
- Provision of GDPR Email Drafts

1 hour
(in person*)

PARTICIPANT INITIATES AND PROGRESSES GDPR REQUESTS WITH TARGETS

4 - 6 weeks

RESEARCHER COACHES AND SUPPORTS THE PARTICIPANT

**INTERVIEW 3**
- Pre-Review Power / Trust Rating
- Describing of GDPR Response / Data
- Detailed Scrutiny of GDPR Response
- Final Power / Trust Rating
- Review of Goals & Hopes (Met?)

2 hours
(video call*)

\* Due to COVID-19, two Interview 2's and all Interview 3's were conducted via Zoom

**Figure 1: A Journey Map of Each Participant's Study Progression**

days for GDPR requests to conclude, a deep dive into each GDPR experience took place. Participants' personal data was not collected by the research team; screen sharing was used to show excerpts to the researcher where the participant wished to do so. Participants were asked to assess the completeness and value of any data returned, and to judge current perceptions of trust and power, whether their hopes had been met, and any general feelings. Answers were recorded in a screen-shared spreadsheet, which was also used to structure the discussion (for a sample cf. attachments).

Questions (asked verbally) began with Yes/No questions, e.g. "Was the observed data returned complete?" which were then explored qualitatively through follow-up discussion as to the reasons for each answer.

Interviews were audio and video recorded, then auto-transcribed using Google Recorder/Zoom, producing a 370,000-word corpus. Transcripts were split up by topic and analysed through reductive coding cycles to produce thematic findings (cf. section 5). Quantitative data from interview spreadsheets was summarised and analysed (see section 4). Sketches, recordings, screenshots and field

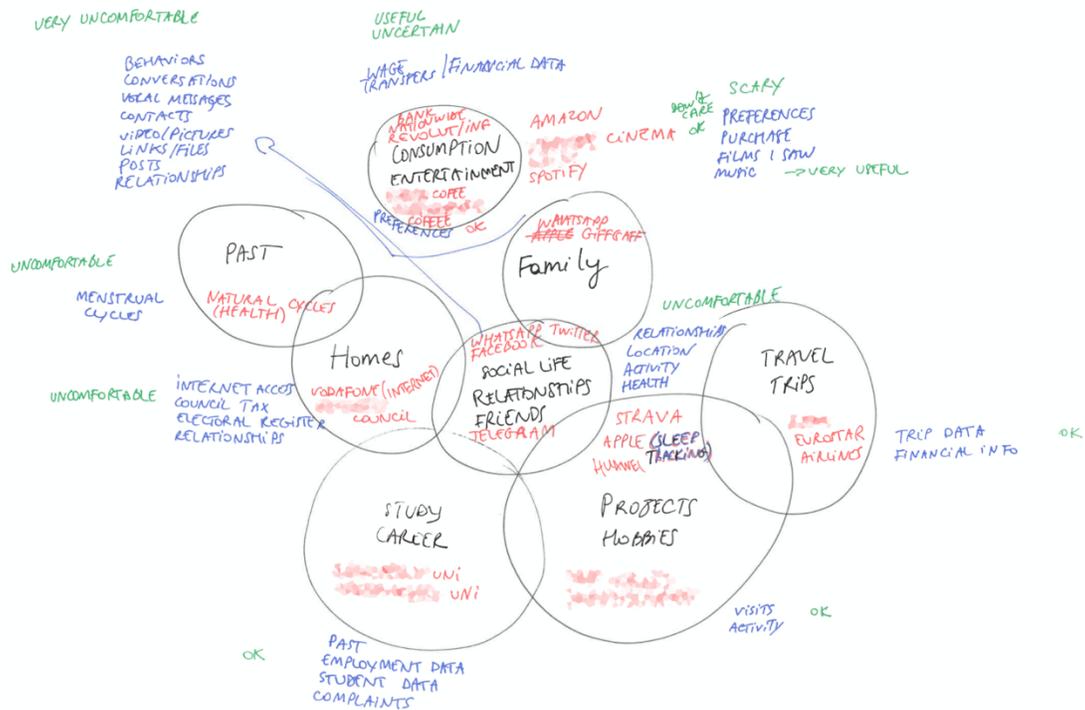

**Figure 2: An Example Life Sketch from Interview 1, with Data Handling Companies in Red, Data Types in Blue, and Feelings in Green**

**Table 1: Types of Data Holding Organisation Targeted for GDPR Requests by Study Participants**

| Type of Company | Company Names[a] |
| --- | --- |
| Major Internet Companies | Apple (3), Amazon, Facebook (4), Google (5) |
| Hardware Companies | Apple (3), Huawei, Google(5), Philips Hue (smart lightbulb manufacturer) |
| Online Platforms/Websites | Airbnb, Bumble (dating site), Check My File, Credit Karma, Direct Line, last.fm, LinkedIn |
| Social Networks & Dating | Facebook (4), Instagram, LinkedIn, Bumble (dating site) |
| Software/App Manufacturers | Freeprints, Niantic (creators of Pokémon Go), Natural Cycles (a menstrual tracker), Revolut, Spotify |
| Transport Companies | Tyne Tunnels, Nexus (Tyne & Wear Metro), LNER |
| Retailers & Loyalty Schemes | Amazon, Tesco, Sainsbury's, Nectar |
| Telcos | Virgin Media, Three |
| Sports Clubs | Sunderland AFC |

[a] Where a company was chosen by more than one participant, number choosing that company is shown in brackets.

notes aided interpretation of the transcripts. Our analysis methodology is detailed in Appendix 3.

## 4 GDPR REQUEST OUTCOMES

### 4.1 Interview 1: GDPR Target Selection

Initially eight participants chose 5 target companies and three chose 4. P9 withdrew from the study due to COVID-19 after Interview 1. Five participants withdrew a chosen company upon further consideration, with reasons including: having one's personal data mixed with other household members (Netflix, Morrisons), not wishing to impact active customer support matters (LNER), and inability to contact the provider by email (see below). One participant selected Newcastle University, which was vetoed by the research team to avoid conflicts of interest. Hence, 41 GDPR subject access requests were made (to 28 distinct data holders; cf. Table 1):

To ensure fairness and consistency, the aim was that all GDPR requests be sent by e-mail to the identified Data Protection Officer, requesting both a subject access request [58] and a data portability request [59] be initiated, and specifically enumerating and asking for those datapoints that the company stated in its privacy policy, as well as those which the GDPR entitles individuals to obtain. To identify these datapoints, company privacy policies were analysed and the necessary information was compiled in personaldata.io's semantic wiki [122], which has a feature to generate bespoke GDPR request emails (cf. attachments). Facebook, Apple, Huawei and Philips Hue offer no contact e-mail address, so shortened email text was pasted into a contact form. TV-sharing website ifun.tv only offered WeChat contact, resulting in the participant (a Chinese citizen) withdrawing ifun.tv due to fear of Chinese government surveillance.

**Table 2: Types of Personal Data Potentially Accessible from Data Holders via GDPR Rights**

| Type of Personal Data | Description | Examples |
|---|---|---|
| Volunteered Data | Data the individual has directly provided to the company through upload, contact or form completion. | Personally Identifiable Information (PII), contact details, user-generated content, photos, files, profiles, settings, communication history, financial information, security credentials, surveys/forms. |
| Observed Data | Data that has been indirectly or automatically collected about the individual through product/service use or customer/staff interaction. | App usage information, behaviour on website, search/browse history, location tracking/tags, activity/health tracking, technical/device information, network/telco/ISP information, cookies & pixel trackers, customer interaction notes. |
| Derived Data | Inferred data or profiles created through algorithmic or human analysis of personal data. | Interest profiles, advertising demographics, market segmentation, customer categorization, product/service recommendations, internal customer codes. |
| Acquired Data | Data obtained or purchased from external sources e.g. civic records, reference agencies, advertisers or third parties. | Public records, information from internet searches, reports or reviews from individuals, electoral roll data, credit checks, fraud checks, criminal checks, e-mail/interest lists from advertisers, information from affiliates, sister companies or partner organisations. |
| Metadata | Information about how the other four categories of data have been handled, including storage, processing, uses, decision-making and external sharing. | Names of third parties data has been shared with, details of where data is stored and when/where it has exited the EU, explanations of how data has been used in automated or human decision making, legal bases for storage and processing. |

As background research, the lead author had conducted over 75 GDPR requests over a two-year period and analysed over 50 companies' privacy policies (and continued this throughout and beyond the interview period). In doing so, he identified different data points and datatypes that organisations possess and common terms used (in policies and in the GDPR). Five overarching types of held personal data were identified, a model which has proved useful and subsequently been adopted internally by research teams at BBC R&D and Hestia.AI. Though it does not use this taxonomy, GDPR grants the right [55, 56] to access all five types of data detailed in Table 2.

## 4.2 Interview 2: Privacy Policy Review and Goal Setting

Participants reviewed and discussed privacy policies for their chosen target companies and were asked to define hopes and expectations for each GDPR request (see Table 4). These most commonly related to *seeing the breadth and depth of data collection* by companies, *understanding what was being inferred* and *how personal data was used*, and to *use such information to better assess trustworthiness* of those companies. Other motivators included the desire to reflect on one's own past data to gain self-insight, and to take control of or delete held data. Minor motivators included learning, creativity, fun, nostalgia, curiosity and the desire to shed light on specific incidents or answer specific questions.

At the conclusion of interview 2, participants were provided with emails and instructions to start their GDPR requests, which progressed as illustrated in Figure 3. Eight requests resulted in no data being obtained, due to either data holder non-responsiveness, inability to access the right account or satisfy ID requirements, or confirmation no data existed. 32 requests (80%) saw at least some data being returned; 10 of these directed the participant to use a publicly-available download dashboard such as Google Takeout. The rest were bespoke deliveries, usually by email (sometimes involving a secured online website to download). Two responses came by post, one as printouts and one on CD-R. 4 of the 22 companies supplying bespoke data packages did not return it within the mandated 30 days. With reference to the five categories in Table 2, participants judged all 32 data returns to be incomplete.

## 4.3 Interview 3: Reviewing the GDPR Response

Upon conclusion or expiry of a participant's GDPR requests, they were invited to discuss GDPR responses in detail. Participants were asked to describe (and optionally show) their received data, then to evaluate the data holder's response for each data type, according to multiple metrics designed to assess the perceived quality of the GDPR request handling and the subjective value of any returned data. Responses were considered quantitatively (cf. Table 3) and qualitatively (cf. section 5).

Table 3 shows quality assessments for each data type, with rows descending by subjective value. Notably, the most valued kinds of data (*derived*, *acquired* and *metadata*) were less frequently returned, especially metadata (returned in 4% of cases). Where data was

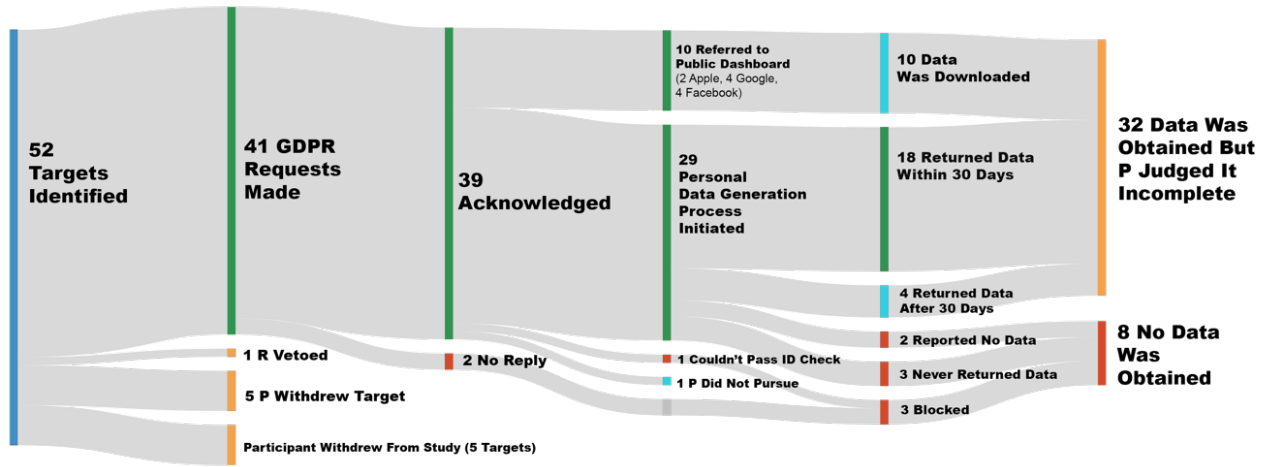

Figure 3: A Sankey diagram giving an overview of the GDPR requests undertaken by our participants (P).

Table 3: Presence and quality assessments of GDPR responses by data category (as percentages)

| Type | Valued? | Returned? | Complete? | Accurate? | Understandable? | Meaningful? | Usable? | Useful? |
|---|---|---|---|---|---|---|---|---|
| Derived | 82% | 39% | 11% | 25% | 40% fully / 40% partially | 40% | 0% | 20% |
| Acquired | 81% | 49% | 19% | 67% | 75% fully / 0% partially | 50% | 25% | 17% |
| Metadata | 73% | 4% | 0% | 0% | 0% fully / 100% partially | 0% | 0% | 0% |
| Volunteered | 57% | 53% | 55% | 92% | 72% fully / 20% partially | 72% | 52% | 58% |
| Observed | 48% | 33% | 20% | 81% | 61% fully / 20% partially | 57% | 52% | 61% |

Percentages represent proportion of "Yes" answers to each question, per data subtype, from all those *where a judgement was given*.
For 'Valued?', participants were asked whether this category of data from each provider would be valuable *if they were to receive it*.

returned in these categories, its quality was poor, often judged as incomplete, inaccurate, unusable and not useful (although *acquired* data was largely understandable). At 53%, even the most returned category, volunteered data, was lacking. Where it was returned, *accuracy* (92%), *meaningfulness* (72%) and *understandability* (72-92%) were high. *Observed data* was least valued and also rarely returned or complete (yet judged of moderate quality). Across all data types, data was only judged complete in 22% of cases, and in 62% of cases personal data specified in privacy policies to be collected was not returned, despite the legal obligation.

When invited to revisit their hopes and anticipated data uses to conclude the third interview (Table 4), participants felt goals were not fully met in 78% of cases, and 54% were not met at all. Specific problem areas included (1) the desire to understand what providers infer from held data (7 participants), unmet in 73% of cases and only fully met in 7% of cases; and (2) the desire to delete one's data, occurring in 10 cases but only met in one. Four wholly unmet hopes were to investigate specific incidents (GDPR responses were often delivered as a one-off package without any backchannel), to secure data, to check accuracy, and to port data to another service.

### 4.4 Perceived Power and Trust

We examined how participants' feelings towards the data holders changed throughout the process. Participants scored trust from 0

(total distrust) to 10 (total trust) and assessed power on a scale of -5 (total provider power) through 0 (balanced power) to +5 (total individual power) then explained their reasons for initial rating and for any change. By repeating the same question at different times, we were able to observe changes in attitude; these changes are summarized in Figure 4. Many participants' attitudes did change as a result of the experience, for both perceived power (45% of cases) and trust (66% of cases). For those with changed attitudes, the change was often negative: in 63% of cases where participants perceived a change in power, that change was a loss in individual power, and in 52% of cases, participants felt more distrustful of GDPR-targeted companies after completing the process (constituting 79% of cases where a trust changed). Privacy policy reviews had distinct impacts from the experience of obtaining and scrutinising returned data (cf. Appendix 2). It is important to note that in some cases GDPR had a positive impact: in 17% of cases participants felt their perceived power had increased, and in 14% of cases participants felt more trusting of providers after GDPR.

## 5 THEMATIC FINDINGS

Here we present outcomes from a deep iterative analysis [80] of transcripts of interviews during the above participant journeys, using a reductive thematic analysis approach (see Appendix 3). We identified three key thematic findings:

**Table 4: Abridged summary of participants' hopes, imagined data uses and goals for GDPR and whether perceived successful**

| Hope or Goal | Distinct instances of this goal | Distinct partici- pants | Specific companies in mind for this goal, if any | Goal perceived as met? | | |
| --- | --- | --- | --- | --- | --- | --- |
| | | | | Unmet? | Partially met? | Fully met? |
| Understand the breadth and depth of what data is collected | 24 | 7 | Amazon, Apple, CheckMyFile, Credit Karma, Facebook, Google, LNER, Nectar, Philips Hue, Spotify, Tesco, Three, Virgin Media | 42% | 17% | 42% |
| Understand what is inferred about you from your data | 15 | 7 | Amazon, Apple, Direct Line, Google, Instagram, last.fm, LNER, Spotify, Tesco, Three | 73% | 20% | 7% |
| Reflect on past activities & gain insights | 14 | 5 | Airbnb, Apple, Google, last.fm, LNER, Tesco, Virgin Media | 57% | 36% | 7% |
| Assess provider trustworthiness | 12 | 6 | Apple, Credit Karma, Direct Line, Facebook, Freeprints, Nectar, Niantic, Sunderland AFC, Tesco, Three | 42% | 42% | 17% |
| Remove your data & control/limit its use | 10 | 3 | Bumble, ifun.tv, Instagram | 90% | 0% | 10% |
| See inside 'black box' algorithms & processes | 9 | 4 | Amazon, Facebook, Google, Tesco | 56% | 11% | 33% |
| Find patterns/habits & track goals | 6 | 5 | last.fm, Nectar, Spotify, Tesco | 17% | 50% | 33% |
| Understand how and why your data is used | 6 | 5 | Direct Line, Google | 50% | 33% | 17% |
| Investigate specific questions or incidents | 4 | 4 | Airbnb, Three, Credit Karma, Instagram | 100% | 0% | 0% |
| Play with, create, hack & remix your data | 3 | 3 | Google | 67% | 0% | 33% |
| Nostalgia, fun & inspiration | 3 | 3 | Spotify, Niantic | 33% | 33% | 33% |
| **OVERALL** | **18 goal types** | **10 people** | – | **54%** | **24%** | **22%** |

For unabridged table, inc. all individual goals/hopes, resultant feelings & grouping by data control vs personal data use, cf. Appendix 1.

1) **Insufficient Transparency**: Organisations appear evasive when responding to GDPR data access requests, leaving people "in the dark" even after making GDPR requests.
2) **Confusing Data**: When presented with their data, people struggle to understand it and relate it to their lives and are not able to make use of it.
3) **Fragile Relationships**: Companies' data practices, in particular their privacy policies and GDPR response handling, can be impactful to customer relationships, carrying a risk of damaging trust but also the potential to improve relations.

These themes are detailed in 5.1, 5.2 and 5.3 respectively.

## 5.1 Many Companies are Evasive and People are "Still in the Dark"

*5.1.1 A Desire for Awareness and Understanding.* In the vast majority (62%) of cases, participants wanted to see, know and understand what data was held about them and how it was used. For example, P11 wanted to know what data was collected by train company LNER when he bought tickets, so he could judge whether it was appropriate:

"I'd be interested to understand what data they have [. . .] Is it just the patterns of my spending on trains, or is it a bunch of other stuff that they're using for advertising to me?"–P11

Beyond volunteered data (see Table 2), what data returns would include was currently unknown to participants. In particular they wanted to gain awareness of what data might have been collected without their knowledge.

"The bit that concerns me is where I don't know what data is being taken by companies. If I'm registering for a library or something, I know [what] data I'm giving to them, but what I don't know is all the other stuff that they're recording"–P9

Participants were equally unaware of inferred data. P4 thought Philips might use smartbulb data to deduce his sleep and TV-watching routines. P7 received targeted pregnancy-related advertisements she *"felt weird about"* because she did not understand why she had been targeted. P5 was concerned that data inferences could affect decision-making, surmising the data holder had greater power than her because *"they're making decisions and [I] don't know how"*. Personal data sharing is insufficiently visible to participants: two participants (P3, P4) targeted GDPR requests to credit-check websites (Credit Karma, CheckMyFile) - P4 wanted to get *"a picture of what other companies can currently expose"*.

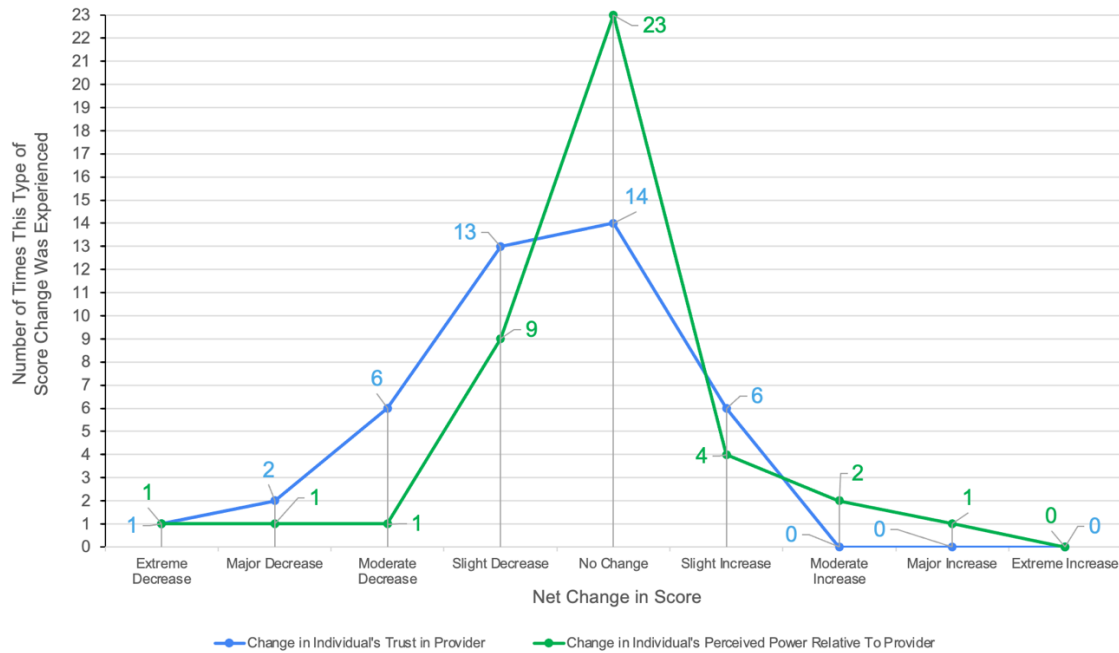

**Figure 4: Distribution of Net Changes in Participant's Perceived Power and Trust Scores over the Study's Duration**

*5.1.2 Non-Compliance without Consequence.* As detailed in 4.2, few requests resulted in timely provision of requested data (44% or 68% depending on whether referral to download portals is excluded or included). Many data holders responded late or not at all—objectively a breach of legislation. However, participants were broadly dissatisfied even when they did receive a GDPR response. In 100% of cases where data was obtained, it was considered incomplete, and this was usually seen as further failure to comply. Participants had reviewed their GDPR rights in Interview 1 (though, as expected [94], most were already aware), so several participants saw this apparent non-compliance relative to their understanding of their rights as a poor quality response, for example:

> "I feel more concerned now, [. . .] what they've given me seemed reasonable. But then comparing against what we asked them for, what I'm legally [entitled to], it's a fraction."—P5

For some, who were sceptical from the start, apparently poor responses were consistent with their expectations; P6 found the incompleteness (14 of 17 subcategories not returned) of Facebook's response *"alarmingly unsurprising"*. Others had expected compliance, and attributed intent to observed incompleteness:

> "I am surprised at Google's unwillingness to provide me with all of the data. . . they haven't provided me with all of my data. And that's not legal."—P7

Many participants, reflecting on a perceived loss of power, felt the prevalence of non-compliance indicated too much power relative to authorities, that regulators apply insufficient pressure and that *"there needs to be more enforcement"* (P11). P6 revised his view of Facebook's power versus his own because he felt that after review he now could clearly see *"which [data] they are prepared to share and which they aren't"*. P11 also framed the selectivity of responses as an exertion of power:

> "It seems like there's a lot of derived data about things like purchases and stuff [that I would expect] that just isn't there. So they're free to not give me the data. That, to me, suggests [despite GDPR] they retain an awful lot of power."—P11

*5.1.3 Inadequate Data Responses.* While in 22% of cases participants did meet their goals (see Table 4), the desire for greater awareness and understanding (cf. 5.1.1) was largely unmet. Only volunteered data such as basic personal information or user-generated content was usually returned complete; this was often viewed as mundane and uninteresting, and the focus on these data types in returns was viewed as evasiveness. Facebook, P6 remarked, *"give you that kind of descriptive boring data which is mainly all publicly available anyway"* and omitted *"the stuff that [is] valuable to them"*.

In general, data returns did not provide the answers participants sought. Many reported *"still"* not knowing what they wanted to find out. P4 said he remained *"in the dark"* (P4). P7 stated that *"even though I did the process correctly, I still didn't get that much back"*. Participants' initially-held concerns remained unaddressed, as with P11:

> "I still am concerned about how much data organisations have, particularly how they link that other data and how data is bought and sold, and I haven't really got any answers on that."

Participants were dissatisfied with the process too; requesting and achieving data access was time-consuming and difficult. *"Jumping through hoops"* was a phrase used independently by four different participants (P4, P5, P7 & P11) to describe the experience. Some found data holders obstructive and unhelpful:

"I feel like they give you a response that [makes it so] you cannot proceed intentionally"–P10

Participants recognised that they had received help and coaching, and felt the processes were so tedious that without that, they may not have persisted. P1 suggested that without the provided template, it would be *"a lot harder to get meaningful data out"*, and P7 attributed her sole successful request to the guidance she had received in progressing it. P5, having experienced problems with expiring links, delayed responses and missed emails, had been surprised at *"how difficult it was just to get my data, and the fact that I had to ask them about six different times"*.

Not all requests were this painful; some were handled smoothly. As P11 put it, *"Some companies make it dead easy to get, but then the data is not massively useful. [. . .] Other companies make it a pain in the neck to get."* Overall GDPR data access was regarded with disappointment. Participants found GDPR ineffective: P10 said *"Frankly, [GDPR] doesn't have as much influence as I expected"* and P1 commented that:

"It's kind of disappointing, because I would have hoped that this process would have levelled the user power versus the organisation power in a way that holds them accountable and [it doesn't] seem to be doing that."

## 5.2 People Struggle to Understand, Use and Control Their Data

*5.2.1 The Search for Personal Value in Data.* Participants anticipated discovering insights about their own lives by browsing and reflecting on their personal data, consistent with personal informatics literature [73]. However, there was a comprehension gap between the useful information they imagined and the actual data returned; data was typically delivered as a bundle of technical files, hard to understand and often delivered without explanation. Some felt (in line with *effective access* [44]) that they lacked the necessary skills or tools to understand the data or use it as *"a non-techie person"* (P11). When the researcher guided P7 to jsonlint.com, an online formatter, she found her JSON-formatted data more understandable. P2 believed data holders must be using tools themselves to make sense of people's data: *"They're not just looking at a JSON file, so I would like to have the same visualisation [as them]."*

By sending people individual data files, data had been removed from the environment in which it has meaning, and divorced it from necessary context for interpretation. This often manifested in the form of internal codes and abbreviations that individuals could not understand. P4 said of his smart-lightbulb data that there was *"so much that it's impossible to know [what it all means]. . . You'd have to spend a few hours going through this and being like, 'OK, what does that line mean, and that symbol, and that code?'"*. This lack of context also materialised as a failure to explain decision-making processes: P5 reflected, when looking at driving scores from Direct Line (which uses a mobile app to monitor her driving), *"I could see the data; it was the score that was weird for me. Like, it doesn't tell you how it's calculated."* P1 noticed that although some companies tried to explain returned data, this varied substantially across providers. He said, *"It would be nice if these companies had a standardised model of how this information is presented to people, so it [could] be easily understood."*

One of the greatest obstacles to understanding that participants faced was large volume of information with no means to quickly digest or navigate it; either very large files, or complex hierarchies of nested directories containing many separate files. Individuals need *summaries* so that they can quickly get a handle on what is—or is not—present. Returned data *"could be valuable if you knew what the hell [was] in there"* (P4). P1 described one of his data responses as *"almost too much [. . .] for a normal person to process and understand."* He said it *"could do with a document detailing, like, 'this is what is in here'"*, and described the disparity across responses as being *"either like death by thirst or death by drowning [. . .] It would be better to drown, but still not ideal"*. Data returns were rarely optimised for understanding.

Our findings shed light on what precisely makes data valuable to individuals: Data was deemed most valuable when it *spans a period of time* and can be related to events in the individual's life over that period. Such data promises novel self-insights for participants.

P2 hoped to see, or be able to construct, breakdowns and charts to examine his food shopping habits. Through GDPR, P10 accessed details of her spending on micro-transactions in Pokémon Go that were not available in the app. P11 hoped to derive insights about his train travel by examining the geography, cost, journey length and patterns of his past journeys through data he hoped to receive from LNER.

Long-duration data carries potential for identifying trends and changes in one's behaviour over time. Such historical datapoints were considered most meaningful, offering a means of remembering, with data potentially serving as a *"window into your past"* (P11). P5 saw value in perusing music-listening data *"just because it's cool to look back on stuff that you've done and you don't necessarily distinctly remember it"*. Generally, the longer period the data covered, the more valuable it was deemed to be; P11 valued last.fm over Spotify because the latter *"only goes back about four or five years"*.

P6 saw data accumulated by service providers as part of a valuable background context to understanding life events in his past: *"I would like to [. . .] build a picture, not just like, 'I remember going to Reykjavik', but if there's other data around that time [I could] sort of paint a biography of myself"*. He described some of his data as *"a kind of personal history that has been quantified and sort of datafied"*.

*5.2.2 Unusable Data Formats.* This potential of returned data to offer personal value adds weight to participants' need to make use of their data. Our participants found returned data formats not only difficult to understand, but difficult to use. Using data meant different things to different participants, with imagined uses including budgeting, record-keeping/archiving, or using the data for creative or fun purposes. Some participants (e.g. P5) saw value in potentially combining data from multiple sources, though this did not turn out to be practical. Participants did not know what data to expect, and generally imagined more useful data than they received.

Participants struggled to interpret and understand returned data enough to identify the useful data or meaningful information they had hoped for. Data formats and response structures were extremely varied. Some reported insufficient machine-readable data. For example, P4 received a Microsoft Word document full of pasted screenshots from an internal portal as part of his response from his ISP Virgin Media, saying its usefulness *"depends on what you want to get out of it, really. If you want to view the data they were about you, it's quite usable. If you want to do something automated [analytical], then it's not"*. P11 found a similar returned screenshot from Tesco's internal systems to be *"completely non-understandable"*. In other cases, the opposite problem occurred, with data being too technical for the participant to use. P10 said of JSON data: *"For normal people who don't understand programming, I feel it's just, there's no use at all."* P7 felt she lacked the technical proficiency to make use of the returned data:

> "They have provided it in formats where I can see that, if I were a developer, I could do things with it, [. . .] but if I was not that sort of person, it might be quite difficult to understand"–P7

P5 saw the potential to use the data but felt a lack of additional explanation or guidance on interpretation:

> "They did give me the data, but not how it fitted together. It's like being given the bricks to a house, and then they're like 'Here's your house'. It doesn't really mean anything when it's just bricks, if you don't know how to put it together."–P5

P11 highlighted a problem with his shopping data that was not just a matter of formatting or skill, but the granularity or focus of the data itself:

> "As a technical person, having a CSV of data is quite useful, potentially, but actually what can I do with that if it's Tesco's internal systems data?"

While 5.2.1 and 5.2.2 and the conflicting demands for both more technical and less technical data might seem contradictory, what we can infer is that participants collectively need *both* usable technical data *and* easy-to-read information summaries - and that those summaries should cover both the relatable life information encoded within the data *and* the information *about* the data, what it means and how to use it; this idea is explored further in [17].

*5.2.3 The Liability of Data You Can't Delete or Control.* Having recognised the potential value of life-data before or during this research, several participants were concerned about personal data holding. P10 said with reference to Bumble: *"Since I found my partner [and therefore no longer need a dating site] I deleted my account and I've been wondering, 'Are they still keeping my data at the back?'"* and expressed a desire to delete her data from both Instagram and Bumble, expecting GDPR to help enforce or verify that deletion, something she could not do otherwise. P8 tolerated the holding of sensitive data only while actively using a service, and thereafter considered it a liability:

> "I now use a different one, but for about a year I used [Natural Cycles'] app to track my menstrual cycle. [It was my] main contraception method, so that's things this company probably has. Now I'm not using it any more, I don't know if they delete the things or not."

Many expressed a desire that data be held only for a short time, and questioned the default practice of data being kept beyond the necessary period:

> "The thing that concerns me is that I haven't used Tesco online for at least four or five years, so why are they hanging on to my IP address from five years ago?"–P11

He went on to spell out the liability he saw in such apparently mundane data being held, which came from from the duration of the data: *"10 years of worth of shopping records. . . how much would that be worth to a health insurance company, and would [Tesco] succumb to the temptation to sell that on?"* P10, a Chinese citizen, identified long-term sources of personal data as an enabler for future privacy violations, saying that *"in China, [there is a trend] that as soon as someone becomes famous, people begin digging [through] their past"*.

Most described the ability to delete or enforce deletion of data as having control over it, and given the current practical lack of such capability felt they had insufficient control. Participants identified how a key first step in gaining control of their data was simply the ability to see it, for accountability, so that they might check the accuracy, security and breadth of collected data and flag any unforeseen concerns. They felt a deeper understanding might lead to an increased sense of individual safety and enable them to make changes in data habits or choice of service provider:

> "I want to understand how much they're keeping. And what they're doing with it. I'm hoping that by knowing that, I might change my behaviour about all the data I accidentally create."–P7

This hope of P7's was unsatisfied, and upon looking back at her experience she remarked:

> "I guess that's one of my criticisms of GDPR in general - that although I can understand what data a company holds about me, there's no obligation for them to tell me what they are doing with it.. And sometimes I think my willingness to give a company data might be quite intrinsically linked with what they're gonna do with it."–P7

In fact, that legal right does exist, but it was not delivered in practice. Participants want to feel aware and in control of their data; this must begin with better data legibility and explanations of data use, accompanied by clear pathways to enable data correction or deletion.

## 5.3 Poor GDPR Handling Can Damage the Fragile Trust Relationship

*5.3.1 Power and Enforced Trust Through Data Holding.* This lack of visibility and control over personal data combined with a sense of being in the dark (cf. 5.1.3) about data practices, caused participants discomfort before, during and after the GPDR process—a sense of facing uncertain risks they felt powerless to change. Discussing their relationships with providers, many expressed emotions ranging from curiosity to anxiety and distrust:

> "I'm curious. . . I wonder what they've got on me. [. . .] If it's anything other than the barest minimum that is necessary for them to do their job [. . .] then I get creeped out by that."–P11

Participants felt most uneasy about the amount of *"intimate"* (P1,P2) data providers collect. P1 was uncomfortable about Facebook having information about social circles. P2 said he felt *"quite vulnerable"* that his Google search terms *"say pretty much everything you have done. . . the most intimate things you were thinking about"*. P11 singled out ISPs as having capability to track everything their customers look at online: *"I don't think you've got much choice about that."*

Some data holders hold so much data that it begins to resemble surveillance: P1, who used *"an absurd amount of [Google's] services"*, reflected that *"if I'm driving, I've got Google Maps open, so they know exactly where I'm going, they know how fast I'm going, they know what I'm listening to while I'm driving. . ."*. Participants saw potential for abuse, fearing such deeply personal knowledge could be *"used against"* them (P2). P11 felt Apple had enough data to *"screw me over"*. P5 felt that Direct Line uses data to *"judge"* her, noting that *"it's not like I can contest the data and say 'Actually, no, I disagree'."* In a more extreme illustration, P10 shared fears that the Chinese government could abuse citizens through data collected by WeChat and Weibu (Chinese services similar to Facebook Messenger and Twitter respectively). Participants were able to identify concrete instances where providers had exploited the personal knowledge they held: in P6's view, Facebook use their knowledge of their users' friendships and relationships to *"hook your attention"* and prevent users deactivating accounts in a *"disingenuous"* manner.

Whether or not data is used nefariously against individuals, thinking about the potential for this caused participants to associate the mass collection of personal data as an acquisition of power over them: *"[Companies that] know a lot about everyone will inherently be able to have power either through persuasion or manipulation"* (P1). P7 saw the **holding of data** as the source of holders' power: *"when I think about other people having my data [. . .] the control isn't sitting with me."*. Others identified the ability of data holders to **deny or limit access to data** as their key source of power:

> "If you're not getting what you perceive to be yours back in completion [sic] then you're not in control of your own data and you have fairly little power over it."–P1

The view of data holders having more power in the service relationship (cf. 2.1) was reflected in participants' evaluations of power balance: this was judged the case in 69% of relationships (rising to 74% after GDPR), whereas in only 17% of cases (unchanged by GDPR) did participants feel they themselves had more power.

Some equated power over a person's data with power over the individual. When asked to define power in the context of data, P8's unprompted comments aligned with our prior findings that data holding serves as a substitute for individual inclusion in decision-making [18]:

> "For me to have power over my data, I think is a fair and normal thing. But for a company to have power

over [my] data means that it's basically a proxy to have power over me."

A key dynamic to understand the value exchange within these relationships is that individuals **sacrifice their data in exchange for value,** that value being the capabilities offered by the services. All 11 participants expressed the idea that they have grown to tolerate *data sacrifice* in exchange for some benefit. P6 tolerates data collection by travel agents because *"they might help me pick a better deal next year."* P11 said he was happy for Tesco to collect data in order to *"profile me to try to sell me more cheese, fine, whatever,"* though expressed caution that he doesn't *"know what else they're doing with it,"* and more generally was *"deeply concerned"* about unseen data trading. The benefit can be convenience too; P10 had logged into Pokémon Go with her Facebook account, knowing that implied data collection by Facebook, *"because it's much easier"*.

Participants often felt data collection was something they had no choice about, and did not like this. Unease over the trade-off being made surfaced most often in the context of recommendations; generally, participants valued data-derived suggestions provided they were *"relevant"* (P1, P8) and not too *"intrusive"* (P1, P6). Data sacrifice is only tolerable up to certain limits: P10 said her sacrifice of data to Niantic was acceptable provided that *"they don't sell where I live or my daily routine"*; however, while their privacy policy promises data is not sold, it does appear location history *is* accessible in some form to third-party advertisers [104]. P8 said relevant music recommendations were *"very useful"* but found Amazon shopping recommendations *"very scary"* because *"I don't want to see that I'm predictable"* and felt that *"if someone out there knows [what I want] before you [it's] like taking agency away from me."*

Permission to collect and use data is knowingly provided by individuals to data holders, but the mechanisms to do so are considered inadequate: P2 felt permission-giving options are *"not granular enough"*. P11 said *"it's not a negotiation at all, it's all or nothing."* Worse, some participants feel permission is coerced from them: P10 observed that Niantic *"pressure you into"* giving continuous access to your location data by tying it to the availability of in-game benefits such that *"you don't want to lose out"*.

Such lack of choice or coercion led to feelings among participants of resignation about data collection: P7 felt that *"it's inevitable that if you want to access services at all [. . .] you automatically have to give [providers] your data."*

Participants felt their data was *"revealing"* (P2,P3,P11) a lot of information about them, and judged their only real option to maintain their privacy was to prevent data collection in the first place by not using that service at all (P1-3, P7, P10, P11), and living with the subsequent lack of service capability; this is Hobson's choice.

### 5.3.2 Perceptions of Data Holders.

Discussions about data-holding service organisations show that factors such as reputation, size and business model were often a major contributor to participants' impressions of companies. For example, P2 described feeling *"more at ease"* with Apple due to their hardware-oriented business model than with Google, who *"make money through data"*. In general, where there is a lack of clarity around how a company makes money, or profit is clearly being made by exploiting sacrificed personal data, there is greater suspicion, while trust is higher in those companies that offer a paid service.

"There [are] no ads. [Natural Cycles is] a paid service, so there's no, like, 'You don't have to pay, but we use your data to make money'." –P8

P2 noted that Apple *"position themselves as a defender of privacy rights"* and along with P11 (another participant who had targeted Apple) favoured them as a result. P10 however had been influenced by a documentary she had seen, becoming suspicious of Apple's control over her hardware.

While attitudes to Apple were generally positive, Facebook– which has, and continues to be, the subject of much negative media attention over its apparently cavalier attitudes towards personal data–was held in much lower regard. P6 said Facebook had *"in every shape or form, shown themselves not to be trusted"*, an opinion formed from *"high-profile news stories where they have done unscrupulous things and are very willing to hand over data"*. P9 reported feeling *"slightly dubious"* about Amazon as a result of *"[press coverage] about their ethics... and the size of them... and the [amount] of data"*. Expectations around data handling influence attitudes toward service providers, though sometimes other factors play a role, such as with P8, who was comforted not just by Natural Cycles' payment model, but the values they project: *"This is woman-empowerment-orientated [sic], so in that sense I think I do [trust them]."*

Beyond perceptions of integrity, participant's direct experiences of interacting with a company affect their feelings toward that provider. P1 found that *"in the same way that Amazon is quite janky [unreliable and awkward to use], Google feels fairly polished and so I trust them more"*. As well as customer/user experience, a perception of receiving value creates trust: P4 said of Google that *"the amount I trust them is in line with the utility I get from them"*. In the context of data sacrifice, high levels of trust do have an effect on customer behaviour:

"When I like the company already, I'm more willing to give them my data." –P2

*5.3.3 Changed Perspectives Through Scrutiny and Transparency.* Longitudinal examination of participant's perceived trust and power across their GDPR experience allows the impact of the experience to be analysed. Trust in data holders tended to diminish through the data request process (cf. Figure 4). Some distrust arose from examination of privacy policies: P5 said Spotify *"shouldn't need to know that much about me, they should just give me music"*. The most noticeable declines in trust occurred between Interview 2 and 3 (when the GDPR request occurred) or within Interview 3 (following examination of returned data), showing that the quality and coverage of the data return *and* the execution of the data request process often have a detrimental effect on trust. Individuals' perceived power, however, did not undergo a corresponding change:

"They've not given me everything back that I thought they'd be collecting, which makes me trust them less. So power-wise, I don't think [anything]'s changed, but trust, I think it has." –P1

The absence or sparsity of derived, acquired and metadata (cf. 4.3) noticeably damaged trust. P1 directly attributed his reduced trust scores to non-compliance (cf. 5.1.2) through failure to return all categories. P5 lowered her Spotify trust score still further upon completion of Interview 3 *"because they didn't say anything about what they're doing with my data or where it's going"*. P8's trust score for Natural Cycles was similarly reduced *"because I think it's hard to get any sensitive data, and it's not made clear what they're using it for"*.

Poor GDPR handling can itself can damage trust, independently of data returned. P2 reduced his Airbnb trust score *"because of the way they've handled [the data request], and the way they've made it hard for me to read the data"*. P7 downgraded her score for LinkedIn *"because I feel like they have my data and [they've] not bothered to find my data, and that makes me feel like I shouldn't trust them quite as much"*.

Participants want greater transparency than current processes deliver, and the failure to foster perceived privacy directly causes distrust:

"If someone's not completely open with you, then you're like, well 'What are you hiding?', which means you trust them less." –P4

Despite impacts on trust, both using GDPR access rights and the wider process of scrutiny and discussion surrounding that process within this study had positive impacts on participants' awareness, offering *"insights into how big companies are actually handling these requests"* (P7) and how to practically use data rights, showing transparency (notwithstanding the inadequacies of current GDPR handling) has an educational benefit. Participants initially expressed wishes to gain insight into data practices to increase accountability and inform decision-making on provider loyalty and privacy settings. GDPR offered the potential to compare data expectations with reality–for example P11 was initially *"curious to find out if [Apple's] marketing claims match their reality around privacy"*. While such broad goals were generally unmet, several found the process thought-provoking and reported feeling more aware about what data they were enabling their providers to gather. P4 felt the process *"got me thinking about, like what other things could I try, and what other sources of personal data are there"*. P8 reflected that *"it's a skill and a kind of knowledge that I think everyone should [have]. I don't think it [should be] normal that I felt so clueless"*. Some commented on the value of understanding GDPR itself through the experience:

"[I] think the exercise was useful in that I understand what a GDPR request can do and what it cannot do. And there's a lot it cannot do. And I think it might seem that it gives you a lot of power, but really, it doesn't." –P2

While considering the negative impacts of the GDPR experience on trust some realised the potential trust-engendering impact that a more transparent response could have brought:

"I think the lack of transparency in a lot of these processes has not helped, you know, if Tesco had [. . .] plain English processes for getting the data and you've got the data in a plain English way, that would do a lot to bolster trust." –P11

In a small number of cases this was witnessed in practice, with a good GDPR response actually increasing participants' trust: P5 reflected that she may have been *"a little harsh"* in her initial judgement of Instagram and said she *"actually really liked what they* *sent. . . in comparison to the others [. . .] I opened Instagram's one and I was like 'this is really cool.'".* P10 was very impressed with the response from Niantic and after GDPR trusted them very highly *"because they replied really fast, the data provided is very detailed, and their attitude towards this whole issue is very positive,"* concluding that they are *"a really nice company"* and even indicating an increased willingness to spend money on their product. P6 trusted Sunderland AFC because *"they were really upfront and I got the data from them first, [. . .] no messing about, the format they gave me just made sense".*

In these comments, we can see an indication that, although the data requests often did not live up to the hopes of the participants, positively engaging with the process is influential and does affect users' outlook. Close attention was paid to the willingness of companies to be transparent and forthcoming, with GDPR representing an opportunity to test organisations on their data practices, thereby assessing their integrity and competence as holders of participants' data.

## 6 DISCUSSION

Our study examined GDPR effectiveness and perceptions in gaining access and control over one's personal data. Participants' experiences support the established power imbalance (cf. 2.1) and suggest GDPR largely fails to empower individuals: both objectively (to the extent possible by our limited sample), in that most companies do not comply fully, and subjectively, in that returned data was often difficult to understand, impractical to use, and raised new questions and concerns. We find indications that swift, transparent, and easy-to-use GDPR procedures can positively affect customer perception of an organisation. In light of these findings, we propose recommendations on how the personal data landscape might be redesigned through policy (6.1) and business practice (6.2), and how individual action can have important impact too (6.3):

### 6.1 Implications for Policymakers: Compliance, Quality and Ongoing Access

Despite significant and obvious investment in dashboards, processes and bespoke data package production, we find (consistent with literature [9]) that poor compliance with the GDPR remains common. Participants' individual experience was overwhelmingly one of disappointment and frustration, with their hopes rarely met. Data holders were not judged to engage meaningfully with the process, and responses typically excluded or obscured data that could have provided the insights sought into organisations' data practices. Evaluations of perceived power compared to data holders remained largely the same or worsened post-GDPR, and confidence in the capabilities of the legislation to shift the balance of power was lacking. Some saw GDPR as a frustrating and time-consuming *"box-ticking exercise"* that did not ultimately help them. While in 7% of cases participants did feel empowered by the GDPR, all participants receiving data were in practice left with additional time-consuming and sometimes technically-skilled work to take advantage of or interpret their returned data. This suggests policymakers need to make changes towards:

1) **Better Compliance Through Enforcement of Complaints.** At present, GDPR enforcement is uneven; each

country has its own DPA (Information Commissioner's Office (ICO) in the UK) and complaints are rarely pursued for individual breaches. Instead, cases are processed by specific DPAs in a form similar to a class action lawsuit. Individuals therefore have little sway when they do raise a complaint. Many GDPR complaints *"become lost or resulted in lengthy delays"* [21] or are erroneously dropped [75]. Until individuals have a clear, effective means to issue complaints [11] that results in enforcement action (or threat of it), it is unlikely individuals can redress their concerns. Data holders must be held accountable if they fail to deliver the full set of data they report possessing, or fail to deliver within the prescribed time frame.

2) **Policies to Enforce Better Quality Responses.** Many participants received data in frustrating formats, including screenshots, printouts or too-technical files littered with acronyms. Data was too technical to understand, and information was not usable (see 5.2), showing demand for *both* human-readable information summaries *and* machine-readable data files. Providers typically return one or the other. Policymakers could suggest data formats or standards for both data and information; helping data portability, improving *effectiveness* [44] and *legibility* [85], and reducing costs by enabling a commons [82] around the building of tools to interpret and understand data to develop. Although standards are emerging [83] due to technological necessity for data unification, without policy incentives adoption lags.

3) **Policies to Enforce Data Access as Ongoing Support, not One-Time Delivery.** A radical redesign of policy is needed to give people the practical outcomes they desire and, according to the GDPR itself, deserve. Data access must be seen as more than "the delivery of data files". People need *understanding* of their data and of its handling to inform their judgements, and this is the measure by which compliance should be assessed. The explanations mandated by GDPR are not forthcoming. Of the 119 hopes expressed in Table 4, 70 (59%) concern greater understanding of data practices. 38 (54%) of these were unmet, and a further 15 (21%) were only partially met. Policies could be more impactful by mandating data holders to support individuals in understanding data and its use, thereby increasing individual agency.

### 6.2 Implications for Data Holders: Earn Trust through Opening Up Data and Enabling Users

While this study and the GDPR itself might seem adversarial to data holders given the goal to reduce their power by imposing new procedures, our findings highlight the importance of personal data to consumer relations. Data holders are likely aware of the paramount role of personal data in decision-making, but may not consider the role of individual perceptions regarding data held about them. Our findings suggest that a failure to satisfy user concerns about the collection and usage of their personal data risks harm to consumer trust and confidence (as well as actual risks from poor handling [20, 77]). Good GDPR handling can increase loyalty, reduce risk and build better relations.

In 52% of cases, our analysis of privacy policies and engaging in GDPR data requests resulted in a decrease in reported trust in the data holder. While such impacts may for now be minimal, as only a small proportion of users read privacy policies [98] and—anecdotal data suggests—an even smaller number conduct GDPR requests, this is likely to change as issues around data privacy and trust continue to take centre stage in global geopolitics [105, 114]. Furthermore, the growing number of businesses focused on "getting your data" or "taking control" [25, 40, 103, 123–125] suggest demand for data access and transparency is increasing. Our findings identify three opportunities for data holders:

1) **Data transparency is an opportunity to increase consumer loyalty and trust.** GDPR rights provide a basis for delivering practical data transparency, enabling organisations to demonstrate they are deserving of trust. By responding clearly and engaging openly and helpfully to access requests, organisations can demonstrate consistency between their promises and actions and demystify the role of users' data in their business model. Research shows explanations can "*ease humans' interactions with technology [. . .], help users understand a system's function, justify system results, and increase their trust*" [42]. In 14% of cases, our participants felt more trusting of the service brand as a result of their GDPR experience (sometimes even displacing prior apprehensiveness or distrust), citing reasons such as speedy, hassle-free responses, clear and understandable data, providers being upfront and open with data, and staff who exhibited a positive attitude to the request.

2) **Data transparency is an opportunity for improved and re-imagined customer relations.** Beyond the opportunity to grow trust, GDPR's data transparency mechanisms can provide individuals with new capabilities for data curation and involvement. By offering users the ability to engage in empowering data interactions, data holders have the opportunity to improve engagement with their organisation. Both parties benefit from viewing user data as a co-owned resource to be curated in partnership with individuals that contribute it. Individuals get a sense of agency, influence and *negotiability* [85]; and service providers, can incentivise users to share more data, increasing data accuracy, reducing liability, and improving ongoing consent approximating dynamic consent [64] rather than today's ineffective and trust-jeopardising models of informed consent [76].

3) **New customer demands indicate untapped business opportunities.** As the 500-member-strong MyData Global organization [87] shows, there is growing demand for personal data empowerment. Personal data is splintered and trapped [1, 16], and people cannot correlate data from different sources in order to reflect, gain insights, and set goals [73]. Due to commercial motivations, service providers generally deliver capabilities within a closed silo, not at the level of an individual's environment [2]. An alternative model places the individual as the point of integration, the centre of their own Personal Data Ecosystem (PDE) [86]. Such life-level capabilities [17] and data involvement [18] have yet to be offered by service providers, and could be a key differentiator. Growing companies such as CitizenMe [126], HestiaLabs [34], Digi.Me [38], Mydex [132], ethi [61], udaptor [103] and exist.io [127] as well as larger organisations like BBC R&D [13] and Microsoft [79] are already starting to innovate in this space.

## 6.3 Implications for Individuals: Becoming Aware of Data's Value and Power, and Demanding More

While disappointment and frustration with GDPR were commonplace, participants gained new understandings; if not of their data, then at least of target companies' approach to access requests. Such new knowledge sufficed to re-affirm or challenge existing attitudes or inform judgements—P1, for example, left Facebook after the study. Even an *attempt* to access data can educate, and even a cursory look at a provider's 'What data do we collect' privacy policy section can provide pause for thought.

Individuals remain in the dark about the collection, use and sharing of their data through a combination of perceived complexity/effort and a lack of evident benefits. Table 4's hopes, alongside PDE's promises (cf. 6.1, 6.2) provide a glimpse of a better future: a world where people take more control of their data and gain actionable self-insights. We propose three insights for individuals:

1) **Your data represents you and defines your user experience.** We hand over data in exchange for access to services, then providers use it to inform product design or recommend content. This 'innocent' handover of data determines how providers control what we can do. Recognizing the crucial role of data (and our limited influence) is the first step towards greater agency and control.

2) **Your data contains meaningful, valuable data about your life.** Data is dry and technical, but participants all sought value within it (cf. 5.2.2). Held data potentially contains rich information about one's life and past activity that may be otherwise inaccessible. This highlights both a risk that others might exploit such insights and a potential benefit that we might harness it ourselves. In this context, data deletion without keeping a copy may be inadvisable. To unlock the trapped value in data, individuals should demand data standards, better access and control mechanisms and insight tools.

3) **Self-education and awareness enable accountability and informed choices.** Our findings highlight a lack of knowledge. Transparency is critical to assessing '*to what extent the bargain is fair*' [69]. GDPR does not always deliver such knowledge, but it remains your right. It cannot be fully refused. By challenging poor GDPR responses and demanding better information, individuals can have impact. Providers are motivated by consumer demand: Download dashboards were created to service at scale. Through persistent public pressure and negative attention, companies can be motivated to improve data access [33]. When exerted relentlessly, GDPR rights can be leveraged to engender small improvements.

## 6.4 Limitations

As with any study, our methodological choices have impacted the findings. Our convenience sample consisted of 11 computer-literate researchers or students, likely to have a better understanding of privacy than an average citizen, limiting generalisability of the findings, but intended to uncover richer insights. Our findings are consistent with—but expand upon—literature: we show that issues with completeness and compliance were reported within the GDPR's first year [9] persist. Challenges in interpreting returned data are known [20], but understood more deeply through our findings (cf. Table 4, Appendix 1).

One objective was to assess companies' GDPR handling. To this end, we trained and guided participants to apply a standard process, allowing us to make fair comparisons. As a consequence, we did not assess participants' skill or ability to utilise their GDPR rights without help; we did not evaluate GDPR 'in the wild'. In service of this objective, participants were taken further on the journey (both in steps taken and in depth of scrutiny) than they might have gone without training and guidance, however we see such education of participants and of the reader as as a social good. We recognise that our coaching influenced participants; nonetheless we were careful to avoid biasing participants' opinions when reviewing privacy policies or examining GDPR returns. All questions were posed from the perspective of (a) the data that providers said they collect and process in their privacy policies and (b) the rights that the GDPR specifies, to ensure discovery of missing data or unfulfilled rights would be considered objectively.

A consequence of our investigation's sociotechnical focus [12] on data access was that we did not examine the impacts of different HDI interface mechanisms. Future research could explore precisely which data presentations, interface features or privacy policy designs have detrimental or positive effects on trust. Best (or worst) practices for GDPR request handling could also be identified.

We provided help in the form of the GDPR request e-mail template, to ensure the right questions were asked and the right terms used, and we advised on how to reply. This helped participants to progress requests where, alone, they might have given up: P1 suggested that without our request wording it would be *"a lot harder to get meaningful data out"*. P7 attributed the success of her requests to our guidance. However, our subsequent experiments with different formats of GDPR request suggest that for broad requests, provided the request is clearly made for a "subject access request AND data portability request", most companies do not respond differently to our detailed request format than a much simpler question; a uniform response is the norm. It is possible the thorough legal language we used (cf. attachments) may have made some smaller companies more fearful and motivated to respond, though non-responsiveness was observed from both small and large companies. Where companies did not respond, we were unable to infer the cause, which might have included anything from disregard to unpreparedness to challenges in complying. It is also important to note that the study took place within the context of the global COVID-19 pandemic, which may affect response times. From our own numerous requests over the last 3 years, we have not seen significant differences in responsiveness since the pandemic began.

## 7 CONCLUSION

We set out to evaluate the human experience of GDPR for service users. Through our qualitative investigation of attitudes to data-centric services before, during and after conducting data access requests, we identify shortcomings in providers' GDPR approaches resulting in unsatisfactory experiences for individuals. When encouraged to draw conclusions on the basis of providers' own promises, individuals' legal rights, and their own hopes (see Table 4), participants (initially lacking awareness of held data and its uses) gained some insights, but were generally still 'in the dark'. We found that providers can seriously damage brand loyalty and trust if comprehensive and well-explained data is not returned in a supportive and open manner (see 5.3). We highlight serious problems with compliance (see 5.1). Participants received data that was incomplete, impractical for use, and they failed to acquire desired explanations. The GDPR does not succeed in its own aim to enhance individuals' rights and control. Participants continued to feel a lack of agency and choice, were largely unable to pursue goals such as data checking, correction or deletion, and their perceived sense of power within the provider relationship was largely unchanged by the experience. Nor does the GDPR allow individuals to adequately pursue their own goals related to accountability, self-reflection or creative data exploration (see 5.2). Individuals cannot acquire power over their data through designing better HDI interfaces alone; this requires redesigned policies and business strategies that take into account the scattered, multi-party sociotechnical context of held data [12, 17]. Our novel contributions include the detailed and rich picture we present of participants' goals for data control and use, supplementing the findings of other works [9, 20, 77] that have identified some of the same issues with GDPR in practice.

From a policy design standpoint, action must be taken to improve both compliance and quality of GDPR responses, but better still would be to provide individuals a more effective and ongoing two-way window into their data (cf. 6.1). The risk of reputational damage should motivate data holders to engage meaningfully with data access requests; such risk can be averted by redesigning both interfaces and processes by seeing data access experiences as an opportunity to build trust and loyalty, perhaps even through establishing progressive co-operative data stewardship relationships (cf. 6.2). While the GDPR experience is often disappointing and frustrating, it can provide insights that help individuals to challenge their assumptions, re-evaluate choices, and in some rare cases, feel empowered to act upon their data. Wider assertion of GDPR rights could demonstrate a desire for data holders to be transparent; without visible demand, little may change (see 6.3).

Our study provides a model for educating individuals about the power of data and conducting future investigations into GDPR experience and effectiveness. SITRA's #digipower investigation [96] (involving our lead author) replicates and expands our study, attempting to improve European data policy by giving elected representatives and public figures across Europe a grounded practical experience of individuals' lack of power over their data.

## ACKNOWLEDGMENTS

Thanks to Paul-Olivier Dehaye and personaldata.io for wiki infrastructure and support with GDPR request generation. Thanks to Tom Maskell, Soheil Human, Euijin Hwang, Ian Forrester, Jasmine Cox, Michael Jelly, Kyle Montague, Sean Peacock, Sunil Rodger, Iain Henderson and the members of the MyData community [87] for valuable discussions that have informed this work.

This work was funded by the EPSRC CDT in Digital Civics (EP/L016176/1). Data supporting this publication is available at https://doi.org/10.25405/data.ncl.c.5759216.v1. Please contact Newcastle Research Data Service at rdm@ncl.ac.uk for further information or access requests.